\documentclass[pre,twocolumn,superscriptaddress,showpacs]{revtex4}
\usepackage{amsmath,amssymb,dsfont,mathtools}
\usepackage[utf8x]{inputenc}
\usepackage[T1]{fontenc}
\usepackage[english]{babel}
\usepackage[dvipdf]{graphicx}

\newcommand{\ddt}{\frac{\mathrm d}{\mathrm {dt}}}
\newcommand{\dds}{\frac{\mathrm d}{\mathrm {ds}}}

\begin{document}
\title{Synchronization in dynamical networks with unconstrained structure switching}
\author{Charo I. \surname{del Genio}}
	\affiliation{School of Life Sciences, University of Warwick, Gibbet Hill Road, Coventry CV4 7AL, UK}
	\affiliation{Max Planck Institute for the Physics of Complex Systems, Nöthnitzer Str. 38, D-01187 Dresden, Germany}
\author{Miguel \surname{Romance}}
\affiliation{Departamento de Mat\'ematica Aplicada, Univ.  Rey Juan Carlos, 28933 M\'ostoles, Madrid, Spain}
\affiliation{Center for Biomedical Technology, Univ. Polit\'ecnica de Madrid, 28223 Pozuelo de Alarc\'on, Madrid, Spain}
\author{Regino \surname{Criado}}
\affiliation{Departamento de Mat\'ematica Aplicada, Univ.  Rey Juan Carlos, 28933 M\'ostoles, Madrid, Spain}
\affiliation{Center for Biomedical Technology, Univ. Polit\'ecnica de Madrid, 28223 Pozuelo de Alarc\'on, Madrid, Spain}
\author{Stefano \surname{Boccaletti}}
	\affiliation{CNR -- Institute of Complex Systems, Via Madonna del Prato, 10, 50019 Sesto Fiorentino (FI), Italy}
	\affiliation{Embassy of Italy in Israel, Trade Tower, 25 Hamered St., 68125 Tel Aviv, Israel}
\date{\today}

\begin{abstract}
We provide a rigorous solution to the problem of constructing
a structural evolution for a network of coupled identical dynamical
units that switches between specified topologies without constraints on
their structure. The evolution of the structure is
determined indirectly, from a carefully built transformation of the
eigenvector matrices of the coupling Laplacians, which are guaranteed
to change smoothly in time. In turn, this allows
to extend the Master Stability Function formalism, which can be used
to assess the stability of a synchronized state. This approach
is independent from the particular topologies that the network
visits, and is not restructed to commuting structures. Also,
it does not depend on the time scale of the evolution, which
can be faster than, comparable to, or even secular with respect
to the the dynamics of the units.
\end{abstract}


\maketitle

Networked structures, in which sets of distributed dynamical systems
interact over a wiring of links with non-trivial topology, are a key
tool to investigate the emergence of collective organization in many
systems of interest~\cite{Alb02,Boc06,Boc14}. The analysis of synchronized
states is particularly relevant, as they play a crucial role in many
natural systems, such as brain networks~\cite{Var01}, or ecological
communities~\cite{Ber99}. In the past decade, the emergence of synchronized
states has been extensively reported and studied~\cite{Boc02}, with
a notable emphasis on the effect of complex static topologies on synchronization
properties~\cite{Pec98,Lag00,Bar02,Nis03,Bel04,Cha05,Hwa05,Mot05,Yao09,Li13,Mas13}.
Nonetheless, static network models do not adequately describe the processes
that arise because of mutations in some biological systems, such as
infectious bacterial population, which are known to have adaptive mutation
rates that can become very highly elevated~\cite{Den06,Csa07}. These
require, instead, the use of time-dependent topologies whose evolution
occurs over time scales that are commensurate with those of the node
dynamics~\cite{And88,Han04}.
A powerful tool to assess the stability of the synchronous solution
in networks of $N$ identical nodes with diffusive coupling, is the so-called
Master Stability Function (MSF)~\cite{Pec98}. If the evolution of such
networks is along structures whose Laplacians commute at all times,
synchronization can be significantly enhanced. In fact, studies have
shown that it can be achieved even when the connection structure at
each time would not allow a synchronized state, in the static case~\cite{Boc06_2,Amr06}.
The far more realistic situation of networks whose coupling matrices
do not necessarily commute has also generated significant interest.
This has led to the development of several results about synchronizability
in systems with specific properties, including the study
of the synchronous state in the case of moving neighbourhood
networks~\cite{Por06}, the rigorous derivation of sufficient
conditions for synchronization in fast switching networks~\cite{Sti06},
and the analysis of the system dynamics in the so-called blinking
limit~\cite{Has13,Has13_2}. In this Article, we propose a framework
that allows to assess the stability of a synchronous solution by means
of an MSF when the evolution of the topology is fully general and unconstrained,
and without assuming any hypothesis on the time scales of the topological evolution.

To this purpose, let us consider a network of $N$ identical systems, evolving according to
\begin{equation}\label{eq1}
\dot{\mathbf{x}}_i = \mathbf{f}\left(\mathbf{x}_i\right) - \sigma \sum_{j=1}^N L_{ij}\left(t\right) \mathbf{h}\left(\mathbf{x}_j\right)\:.
\end{equation}
Here dot denotes time derivative, $\mathbf{x}_i$ is an $m$-dimensional
row vector describing the state of the $i^\mathrm{th}$ node, $\sigma$
is the interaction strength, and $\mathbf f$ and $\mathbf h$ are two vectorial
functions describing the local dynamics of the nodes and the output from
a node to another, respectively. Also, $L\left(t\right)=S\left(t\right)-W\left(t\right)$
is the Laplacian of the network describing the time evolution of the connections.
In this expression, $W$ is the weighted adjacency matrix, and $S$ is the
diagonal strength matrix of the network: $S_{ij}=\delta_{i,j}\sum_{k=1}^N W_{ik}$.
To fix the ideas, assume that the network has an initial structure, with
Laplacian $L_0$, that is constant from time $t=0$ to $t=t_0$. Also assume
that, at time $t_1>t_0$, the network is found in a configuration with
Laplacian $L_1≠L_0$. Note that in the following we consider Laplacians
with Hermitian structure. Also, as $L\left(t\right)$ is a Laplacian matrix,
the sum of each row vanishes, its diagonal elements are strictly positive
and its off-diagonal elements are non-positive. Now, let $\mathbf{x_s}$
be the state vector indicating the synchronized solution, and define the
$mN$-dimensional column vector $\delta\mathbf X=\left(\delta\mathbf x_1,\dotsc,\delta\mathbf x_N\right)^\mathrm{T}$,
representing the global deviation from the synchronized state. From Eq.~\ref{eq1},
to linear order in $\delta\mathbf X$, one has
\begin{equation}\label{eq:var}
 \delta\dot{\mathbf X} = \left(\mathds 1\otimes\mathrm J
\mathbf f\left(\mathbf{x_s}\right) - \sigma L\left(t\right)\otimes\mathrm J \mathbf h
\left(\mathbf{x_s}\right)\right)\delta\mathbf X\:,
\end{equation}
where $\mathds 1$ is the $N$-dimensional identity
matrix, $\otimes$ denotes the direct product, and
$\mathrm J$ is the Jacobian operator. The vector
$\delta\mathbf X$ can be written at each time as
a sum of direct products of the eigenvectors $\mathbf v_i$
of $L\left(t\right)$ and a time-dependent set of
$N$ $m$-dimensional row-vectors $\boldsymbol \eta_i\left(t\right)$:
\begin{equation*}
 \delta\mathbf X=\sum_{i=1}^N\mathbf v_i\left(t\right)\otimes\boldsymbol \eta_i\left(t\right)\:.
\end{equation*}
Multiplying $\mathbf v_j^{\mathrm T}$ from the
left to both sides of Eq.~\ref{eq:var}, one gets
\begin{equation}\label{variation}
 \ddt\boldsymbol \eta_j=\mathbf K_j\boldsymbol \eta_j\ \ \boxed{-\sum_{i=1}^N\mathbf v_j^{\mathrm T}
\left(t\right)\cdot\ddt\mathbf v_i\left(t\right)\boldsymbol\eta_i}\:,
\end{equation}
where, for the sake of brevity, we defined
$\mathbf K_j\equiv\left(\mathrm J \mathbf f\left(\mathbf{x_s}\right) - \nu_j\mathrm J \mathbf h(\mathbf{x_s})\right)$
and $\nu_j\equiv\sigma\lambda_j\left(t\right)$, in which $\lambda_j\left(t\right)$
is the $j^\mathrm{th}$ eigenvalue of $L\left(t\right)$.
To assess the stability of this dynamical system, one can compute the
Master Stability Function, which represents the dependence on $\nu$ of
the largest Lyapunov exponent $\Lambda_{\max}$ associated to the equations~\ref{variation}.
Then, the stability criterion for a given $\nu$ is that the time averages
of $\Lambda_{\max}$ in the direction of all eigenvectors are negative.
This allows to study systems with highly non-trivial behaviour. As an
example, one can consider an evolving system where each ``frozen'' connection
topology has at least one direction in which $\Lambda_{\max}$ is positive,
but the synchronization manifold is transversely stable~\cite{Boc06_2}.
Similarly, one can detect instabilities introduced by the evolving nature
of the Laplacian in systems where the synchronization manifold in each
frozen configuration is attracting~\cite{Jos08}. This is particularly
useful, as it is well-known that non-linear perturbations in a system
can destroy the stability of the synchronized state~\cite{Bar02_2}.

Note that using the MSF is not the only possible
method to assess the stability of a synchronized
state. For instance, one could construct the Lyapunov
function for the synchronization manifold. This
would guarantee the stability since it is a necessary
and sufficient condition, while in the general case
the MSF is only a necessary one. However, while
it is certainly possible to build the Lyapunov function
in some specific cases~\cite{Wu07}, a general construction
method is not known. Also, it should be noted that
in the absence of fixed points or other attracting
sets far from the synchronization manifold, the
MSF provides a sufficient stability condition as
well, thereby becoming a widely used approach~\cite{Sor11}.

To use the MSF method, one must first note
that the boxed term in Eqs.~\ref{variation}
explicitly depends on the time variation of
the eigenvectors of $L$. If the Laplacians
$L_0$ and $L_1$ commute, one can choose to
study the problem in the common basis of eigenvectors.
In this reference frame the eigenvectors do
not change. Thus, the boxed term vanishes
and Eqs.~\ref{variation} reduce to a set of
variational equations. However, if we allow
the network to switch to a different structure
without imposing the extra requirement of
commutativity, the eigenvector variation must
be taken into account. Note that this forbids
instantaneous jumps between the two structures,
as a sudden change in the eigenvectors would
cause their derivatives to be not defined.
Therefore, the goal is constructing a smooth
evolution process from $t=t_0$ to $t=t_1$
to allow the system to evolve between between
the two topologies while keeping the eigenvector
elements differentiable. To achieve this,
we first consider the matrices $A$ and $B$,
consisting of the eigenvectors of $L_0$ and
$L_1$, respectively, and describe how to transform
one into the other via a proper rotation around
a fixed axis. Then, we use this framework
to find a transformation between $L_0$ and
$L_1$. The rotation evolving $A$ into $B$
takes the form of a one-parameter transformation
group $G_s$ such that $G_0A=A$ and $G_1A=B$.
Note that, as we will use this mapping to
build the Laplacian, we must also impose the
extra requirement that the vector $\mathbf a$,
corresponding to the null eigenvalue, is kept
constant by $G_s$ for all $0\leqslant s\leqslant 1$.

In general, the transformation
$O$ from $A$ to $B$ is a rotation,
which can be found by solving
the linear system of $N^2$ equations
in $N^2$ variables $OA=B$. It
is convenient to work in the basis
defined by the matrix $A$. In
this basis, $A\equiv\mathds{1}$.
Without loss of generality, assume
that the conserved vector $\mathbf a$
is the first vector of $A$. Then,
the transformation matrix $O$
has the form
\begin{equation}\label{tmat} O=
\begin{pmatrix}
 1      & 0      & 0      & \dotsm & 0\\
 0      & O_{22} & O_{23} & \dotsm & O_{2N}\\
 0      & O_{32} & O_{33} & \dotsm & O_{3N}\\
 \vdots & \vdots & \vdots & \ddots & \vdots\\
 0      & O_{N2} & O_{N3} & \dotsm & O_{NN}
\end{pmatrix}\:.
\end{equation}

As $O\in O\left(N\right)$, it is a proper rotation
if its determinant is~1, or a composition of rotations
and reflections if its determinant is $-1$. The determinant
of $O$ equals the determinant of the minor $O'$ obtained
by removing the first row and the first column from
$O$. Thus, we only need to find a solution to the
problem in $N-1$ dimensions. We will henceforth use
primes when referring to objects in $N-1$ dimensions.

From the considerations above, it is $G'_0=\mathds{1}'$
$G'_1=O'$, and, of course, $O'\in O\left(N-1\right)$. Thus,
the problem is equivalent to determining the possibility
of finding a path between the identity and $O'$. If $\left|O'\right|=1$,
then $O'\in SO\left(N-1\right)$. But $SO\left(N-1\right)$
is the connected identity component of the orthogonal group,
and additionally, since it is a manifold, it is path-connected~\cite{War71}.
Thus, for every orthogonal $\left(N-1\right)\times\left(N-1\right)$
matrix in it, there is a continuum of orthogonal matrices
of the same dimension connecting it to the identity. Each
point along this path corresponds to an orthogonal matrix
that can be embedded in $SO\left(N\right)$ by adding a
$1$ in the top left corner. Since every such embedded matrix
keeps the synchronization manifold vector invariant, a
parametrization of the path provides a direct solution
to the original problem.

If, instead, $\left|O'\right|=-1$, then $O'\in O\left(N-1\right)\setminus SO\left(N-1\right)$.
While $O\left(N-1\right)\setminus SO\left(N-1\right)$ is also a connected topological space, the
identity does not belong to it~\cite{War71}. Thus, no path connects the identity to $O'$. However,
in our case, the labeling of the vectors is irrelevant. In other words, provided that the vector
$\mathbf a$ is kept constant, one can arbitrarily swap two vectors in the basis given by the matrix
$A$, obtaining a new matrix $C$. Of course, this imposes a swap of the corresponding columns in
the transformation matrix $O$ as well. But swapping two rows in a matrix changes the sign of its
determinant. This means that the new matrix $O'\in SO\left(N-1\right)$, and a path connecting
it to the identity can be found. The only consequence of the swap is a change of the order of
the eigenvalues. As we are considering unlabelled networks, the problem can always be solved.
To build an explicit solution, we factor the transformation $O'$ into rotations and reflections
in mutually orthogonal subspaces.

Before describing the actual procedure, we recall a useful, twofold result.
First, any orthogonal operator $X$ in a normed space over $\mathbb{R}$ induces
a 1-\ or 2-dimensional invariant subspace. To find one such subspace, first
define the unitary operator $U$ acting on $\mathbf a+\mathrm i\mathbf b$ as
$U\left(\mathbf a+\mathrm i\mathbf b\right)=X\mathbf a+\mathrm iX\mathbf b$,
where $\mathbf a$ and $\mathbf b$ belong to $\mathbb{R}^N$. Then, find a non-vanishing
eigenvector $\mathbf x=\mathbf {x_R}+\mathrm i\mathbf{x_I}$ of $U$. The span
of $\mathbf{x_R}$ and $\mathbf{x_I}$ defines the invariant subspace: applying
$X$ to any linear combination of $\mathbf{x_R}$ and $\mathbf{x_I}$ produces
a vector that is still a linear combination of $\mathbf{x_R}$ and $\mathbf{x_I}$.
Also, if the corresponding eigenvalue is complex, $\mathbf{x_R}$ and $\mathbf{x_I}$
are orthogonal.

Using this, we can describe the following algorithmic procedure
to build a transformation $O$ of $A$ into $B$:
\begin{enumerate}
 \item Express the problem in the basis $A$, in which $O$ has the form of Eq.~\ref{tmat}.
 \item Consider the operator $O'$ obtained from $O$ by removing the first row and the first column, and let $d$ be its dimension.
 \item Build the operator $U$ that acts on $\mathbf a+\mathrm i\mathbf b$ as
$U\left(\mathbf a+\mathrm i\mathbf b\right)=O'\mathbf a+\mathrm iO'\mathbf b$,
where $\mathbf a$ and $\mathbf b$ belong to $\mathbb{R}^d$.
 \item Find an eigenvector $\mathbf x=\mathbf{x_R}+\mathrm i\mathbf{x_I}$ of $U$, with eigenvalue $\lambda$.
 \item Normalize $\mathbf{x_R}$ and $\mathbf{x_I}$.
 \item If $\lambda\in\mathbb{R}$ then
\begin{enumerate}
  \item Pick the non-vanishing component between $\mathbf{x_R}$ and $\mathbf{x_I}$.
If both are non-zero, choose one randomly. Without loss of generality, assume this is $\mathbf{x_R}$.
  \item Create $d-1$ other orthonormal vectors, all orthogonal to $\mathbf{x_R}$,
and arrange all these vectors so that $\mathbf{x_R}$ is the last of them. This set
of vectors is an orthonormal basis $C$ of $\mathbb{R}^d$.
  \item Change the basis of the $d$-dimensional sub-problem to $C$. In this basis,
all the elements in the last row and in the last column of $O'$ will be 0, except
the last one, which will be $\pm 1$.
  \item If $d>1$, consider a new operator $O'$ obtained from the old $O'$ by removing the last
row and the last column. Let $d$ be its dimension, and restart from step~3. Otherwise stop.
\end{enumerate}
If instead $\lambda\notin\mathbb{R}$, then
\begin{enumerate}
  \item Create $d-2$ other orthonormal vectors, all orthogonal to
$\mathbf{x_R}$ and $\mathbf{x_I}$, and arrange all these vectors
so that $\mathbf{x_R}$ and $\mathbf{x_I}$ are the first two of them.
This set of vectors is an orthonormal basis $C$ of $\mathbb{R}^d$.
  \item Change the basis of the $d$-dimensional sub-problem to $C$.
In this basis, all the elements in the first two rows and in the first
two columns of $O'$ will be 0, except the first two.
  \item If $d>2$, consider a new operator $O'$ obtained
from the old $O'$ by removing the first two rows and the
first two columns. Let $d$ be its dimension, and restart
from step~3. Otherwise stop.
\end{enumerate}
\end{enumerate}

At each iteration of the steps above, 1~or 2~dimensions
are eliminated from the problem. All the subsequent changes
of base leave the already determined elements of $O$ unchanged,
because they act on orthogonal subspaces to those already
eliminated. The procedure reconstructs $O$ piece by piece
with a block-diagonal form, in which the blocks correspond
to the action of the orthogonal operator on the invariant
subspaces. If the subspace is 1-dimensional, then the block
is a single $\pm1$ element. If instead the subspace is 2-dimensional,
then its block is either a rotation or a reflection, i.e.,
it is either $
\begin{pmatrix}
 \cos\alpha & -\sin\alpha\\
 \sin\alpha & \cos\alpha\\
\end{pmatrix}$
or $
\begin{pmatrix}
 \pm 1 & 0\\
 0     & \mp 1\\
\end{pmatrix}$.
Thus, the 1-dimensional invariant subspaces
induced by the operator correspond either to
leaving one direction untouched, or to reflecting
the system about that direction. Conversely,
the 2-dimensional invariant subspaces correspond
to rotations in mutually orthogonal planes.

Once this form of $O$ is found, permute the basis vectors
from the second onwards so they correspond, in order, first
to all the actual rotation blocks, then to the $-1$ elements,
and finally to the $+1$ elements. The new form of the transformation
matrix, $O_N$, is simply $O_N=TOT^{-1}$, where $T$ is the
required change-of-basis matrix.

Note that the determinant of $O_N$ could still be~$-1$.
However, as seen before, in this case, one can relabel
two vectors of the original basis, inducing a swap of the
corresponding columns of $O_N$. To perform this, first
note that, if $\left|O_N\right|=-1$, then the number of
$-1$ elements in $O_N$ must be odd. Then, there are three
possible cases.

If $O'_N$ has at least one $+1$ element,
swap the basis vectors corresponding to
the first $-1$ and the first $+1$ elements
in $O'$. Then, the first block after the
``$\sin$--$\cos$'' blocks is $\begin{pmatrix} 1&0\\0&-1\end{pmatrix}$.
Swapping the labels of the the two corresponding
vectors, makes the block in $O'$ become $
 \begin{pmatrix}
  0 & 1\\
  -1 & 0
 \end{pmatrix}
=\begin{pmatrix}
  \cos\left(-\frac{\pi}{2}\right) & -\sin\left(-\frac{\pi}{2}\right)\\
  \sin\left(-\frac{\pi}{2}\right) & \cos\left(-\frac{\pi}{2}\right)
 \end{pmatrix}$.

If instead $O'_N$ has no $+1$ elements and only one $-1$ element,
the basis vectors to swap are those corresponding to the first two
vectors in the last $3\times 3$ block of $O'$, that becomes
\begin{equation*}
 M = \begin{pmatrix}
      -\sin\vartheta_k & \cos\vartheta_k & 0\\
      \cos\vartheta_k  & \sin\vartheta_k & 0\\
      0 & 0 & -1
     \end{pmatrix}.
\end{equation*}
Now, perform one more basis change, leaving
all the basis vectors unchanged, but mapping
the last three into the eigenvectors of $M$,
which are
\begin{equation*}
 V = \begin{psmallmatrix}
      0 & \left(-\sec\vartheta_k-\tan\vartheta_k\right)\sin\left(\frac{\pi}{4}-\frac{\vartheta_k}{2}\right) & \left(\sec\vartheta_k-\tan\vartheta_k\right)\sin\left(\frac{\vartheta_k}{2}+\frac{\pi}{4}\right)\\
      0 & 1 & 1\\
      1 & 0 & 0
     \end{psmallmatrix}\:.
\end{equation*}
The new form of $M$ becomes
\begin{equation*}
 M' = V^\mathrm{T}MV = \begin{pmatrix}
                        -1 & 0  & 0\\
			0  & -1 & 0\\
			0  & 0  & 1
                       \end{pmatrix}.
\end{equation*}

Finally, if $O'_N$ has no $+1$ elements and at least three $-1$ elements,
swap the basis vectors corresponding to the first two after the ``$\sin$--$\cos$''
blocks. Their $3\times 3$ block is now
\begin{equation*}
M = \begin{pmatrix}
  0  & -1 & 0\\
  -1 & 0  & 0\\
  0  & 0  & -1
 \end{pmatrix}.
\end{equation*}
Next, do a change of basis as described in the previous case.
The eigenvector matrix of $M$ is
\begin{equation*}
 V = \begin{pmatrix}
      0 & \frac{1}{\sqrt{2}} & -\frac{1}{\sqrt{2}}\\
      0 & \frac{1}{\sqrt{2}} & \frac{1}{\sqrt{2}}\\
      1 & 0 & 0
     \end{pmatrix}\:,
\end{equation*}
so, after the basis change, the new form of $M$ is once more
\begin{equation*}
 M' = V^\mathrm{T}MV = \begin{pmatrix}
                        -1 & 0  & 0\\
			0  & -1 & 0\\
			0  & 0  & 1
                       \end{pmatrix}.
\end{equation*}

Regardless of the original value of $\left|O_N\right|$,
take now every two subsequent $-1$ elements, if there are
any, and change their diagonal block into $
\begin{pmatrix}
 \cos\pi & -\sin\pi\\
 \sin\pi & \cos\pi\\
\end{pmatrix}$.
This yields the final general form for the transformation matrix,
which can be turned into the required transformation group via the
introduction of a parameter $s\in\left[0,1\right]$:
\begin{widetext}
\begin{equation}
G_s=
\begin{psmallmatrix}
 1 & 0 & 0 & \dotsm & 0 & 0 & 0 & 0 & 0 & 0 & 0 & \dotsm & 0 & 0\\
 0 & \cos\left(\vartheta_1s\right) & -\sin\left(\vartheta_1s\right) & \dotsm & 0 & 0 & 0 & 0 & 0 & 0 & 0 & \dotsm & 0 & 0\\
 0 & \sin\left(\vartheta_1s\right) & \cos\left(\vartheta_1s\right) & \dotsm & 0 & 0 & 0 & 0 & 0 & 0 & 0 & \dotsm & 0 & 0\\
 \scriptscriptstyle \vdots & \scriptscriptstyle \vdots & \scriptscriptstyle \vdots & \scriptscriptstyle \ddots & 0 & 0 & 0 & 0 & 0 & 0 & 0 & \dotsm & 0 & 0\\
 0 & 0 & 0 & 0 & \cos\left(\vartheta_ks\right) & -\sin\left(\vartheta_ks\right) & 0 & 0 & 0 & 0 & 0 & \dotsm & 0 & 0\\
 0 & 0 & 0 & 0 & \sin\left(\vartheta_ks\right) & \cos\left(\vartheta_ks\right) & 0 & 0 & 0 & 0 & 0 & \dotsm & 0 & 0\\
 0 & 0 & 0 & 0 & 0 & 0 & \cos\left(-\frac{\pi}{2}s\right) & -\sin\left(-\frac{\pi}{2}s\right) & 0 & 0 & 0 & \dotsm & 0 & 0\\
 0 & 0 & 0 & 0 & 0 & 0 & \sin\left(-\frac{\pi}{2}s\right) & \cos\left(-\frac{\pi}{2}s\right) & 0 & 0 & 0 & \dotsm & 0 & 0\\
 0 & 0 & 0 & 0 & 0 & 0 & 0 & 0 & \cos\left(\pi s\right) & -\sin\left(\pi s\right) & 0 & \dotsm & 0 & 0\\
 0 & 0 & 0 & 0 & 0 & 0 & 0 & 0 & \sin\left(\pi s\right) & \cos\left(\pi s\right) & 0 & \dotsm & 0 & 0\\
 \scriptscriptstyle \vdots & \scriptscriptstyle \vdots & \scriptscriptstyle \vdots & \scriptscriptstyle \vdots & \scriptscriptstyle \vdots & \scriptscriptstyle \vdots & \scriptscriptstyle \vdots & \scriptscriptstyle \vdots & \scriptscriptstyle \vdots & \scriptscriptstyle \vdots & \scriptscriptstyle \vdots & \scriptscriptstyle \ddots & \scriptscriptstyle \vdots & \scriptscriptstyle \vdots\\
 0 & 0 & 0 & 0 & 0 & 0 & 0 & 0 & 0 & 0 & 0 & \dotsm & 1 & 0\\
 0 & 0 & 0 & 0 & 0 & 0 & 0 & 0 & 0 & 0 & 0 & \dotsm & 0 & 1\\
\end{psmallmatrix}\:.
\end{equation}
\end{widetext}

Notice that when $s=0$, $G_s=G_0=\mathds{1}$, and when $s=1$, $G_s=G_1=O$.
Also, for every value of $s$, the determinant of $G_s$ is always 1, and the
first vector is kept constant, which means that $G_s$ describes a proper rotation
around the axis defined by the eigenvector corresponding to the synchronization
manifold. Moreover, the first vector has always been left untouched by all
the possible basis transformations. Thus, as $s$ is varied continuously between~0
and~1, the application of $G_s$ sends $A$ into $B$ continuously, as needed.

Finally, to describe how to obtain a transformation
between the two Laplacians $L_0$ and $L_1$, let $R$
be the change-of-basis matrix resulting from the composition
of all basis changes done in step~6 of the method,
the permutation of the vectors to obtain $O$, and the
possible final adjustment in case of negative determinant.
Also, to simplify the formalism, in the following we
let $B_0\equiv A$ and $B_1\equiv B$. Since $B_0$ and
$B_1$ are matrices of eigenvectors, it is $B_0^{\mathrm T}L_0B_0 = D_0$
and $B_1^{\mathrm T}L_1B_1 = D_1$, where $D_0$ and
$D_1$ are the diagonal matrices of the eigenvalues
of $L_0$ and $L_1$. Also, as $G_s$ is a group of orthogonal
transformations, it is $ G_sRB_0=RB_s\quad\forall0\leqslant s\leqslant 1$,
where $B_s$ is a basis of $\mathbb{R}^N$. Multiplying
this by $R^{\mathrm T}$ on the left yields
\begin{equation}\label{BTdef}
 R^{\mathrm T}G_sRB_0 = R^{\mathrm T}RB_s = B_s\:,
\end{equation}
where we used $R^{\mathrm T}=R^{-1}$.

Now, note that all the basis changes performed
are between orthonormal bases. Thus $R$ is an
isometry, as is any $G_s$, since they are all
proper rigid rotations, and any $B_s$ defines
an orthonormal basis of $\mathbb{R}^N$. Then,
the Laplacian for the parameter $s$ is given by
the matrix $L_s$ that solves the equation
\begin{equation}\label{BTDT}
 B_s^{\mathrm T}L_sB_s=D_s\:,
\end{equation}
where $D_s$ is a diagonal matrix whose
elements are the eigenvalues of $L_s$,
and $B_s$ consists of the eigenvectors
of $L_s$. However, the equation above
has two unknowns, namely $L_s$ and $D_s$.

This provides a certain freedom in describing
the evolution of the eigenvalues of the Laplacians.
For instance, one can choose the simplest evolution,
which is given by a set of linear transformations.
Then, for all $1\leqslant i\leqslant N$,
\begin{equation}\label{valueevol}
 \lambda_i^{(s)} = \left(1-s\right)\lambda_i^{(0)}+s\lambda_i^{(1)}\:,
\end{equation}
where $\lambda_i^{(0)}$ and $\lambda_i^{(1)}$
are the $i^\mathrm{th}$ eigenvalues of $L_0$
and $L_1$, respectively. Note that this allows
for the possibility of degeneracy of some eigenvalues
for some particular value of the parameter $s^\ast$.
However, the transformation we described leaves
all the eigenvectors distinct and separate throughout
the evolution. Thus, the Laplacian can always
be diagonalized for any value of $s$. Multiplying
Eq.~\ref{BTDT} on the right by $B_s^{\mathrm T}$
yields $B_s^{\mathrm T}L_sB_sB_s^{\mathrm T}=D_sB_s^{\mathrm T}$,
hence $B_s^{\mathrm T}L_s=D_sB_s^{\mathrm T}$.
Multiplying this on the left by $B_s$, it is
$B_sB_s^{\mathrm T}L_s=B_sD_sB_s^{\mathrm T}$,
hence $L_s=B_sD_sB_s^{\mathrm T}$. Substituting
Eq.~\ref{BTdef} into this last equation gives
\begin{equation}\label{LvsD}
\begin{split}
 L_s = R^\mathrm{T}G_sRB_0D_s\left(R^\mathrm{T}G_sRB_0\right)^\mathrm{T}\\=R^\mathrm{T}G_sRB_0D_sB_0^\mathrm{T}R^\mathrm{T}G_s^\mathrm{T}R\:.
\end{split}
\end{equation}

In the equation above, $B_0$ is known, $R$ and $G_s$ have been explicitly
built, and $D_s$ is completely determined by Eq.~\ref{valueevol}. Thus, Eq.~\ref{LvsD}
defines the Laplacian for any given value of the parameter $s$. The evolution
of the eigenvalues is imposed by Eq.~\ref{valueevol},
and the evolution of the eigenvectors is given by Eq.~\ref{BTdef}.

It is important to stress that the solution
given by the linear evolution of the eigenvalues is not
unique. There could be, in principle, many allowed transformations
of $L_0$ into $L_1$, each characterized by a specific eigenvalue
evolution. However, the difference between solutions is
only in the $\mathbf K_j$ term in Eq.~\ref{variation},
since the boxed term does not depend on the eigenvalues.
Thus, the specific choice of eigenvalue evolution could
change the phenomenology of the system studied, but would
not modify how the switching between general structures
affects the stability. This is akin to the results presented
in Ref.~\cite{Boc06_2}, which can be considered a special
case of the present treatment that occurs when all the
Laplacians commute.

With this result, one can finally describe the system evolution
through unconstrained topologies. From $t=0$ to $t=t_0$, the boxed
term in Eq.~\ref{variation} vanishes. To compute its value during
the switch, first note that the $i^\mathrm{th}$ eigenvector at
time $t$ is the $i^\mathrm{th}$ column of $B_s$, with $s\equiv\frac{t-t_0}{t_1-t_0}$.
But then, using Eq.~\ref{BTdef}, the $k^\mathrm{th}$ element of
the $i^\mathrm{th}$ eigenvector is
\begin{equation*}
\begin{split}
 \left(\mathbf{v}_i\right)_k = \left(B_s\right)_{ki} = \left(R^\mathrm{T}G_sRB_0\right)_{ki} = \\ = \sum_{r=1}^N\sum_{q=1}^N\sum_{x=1}^N R_{rk}\left(G_s\right)_{rq}R_{qx}\left(B_0\right)_{xi}\:.
 \end{split}
\end{equation*}

Notice that in the equation above the only term that depends
on time is $\left(G_s\right)_{rq}$, since $R$ is just a change-of-basis
matrix, and $B_0$ is the matrix of eigenvectors of $L_0$ at
time $t=0$. Therefore, it is
\begin{equation*}
 \ddt\left(\mathbf{v}_i\right)_k = \sum_{r=1}^N\sum_{q=1}^N\sum_{x=1}^N R_{rk}R_{qx}\left(B_0\right)_{xi}\frac{1}{t_1-t_0}\dds\left(G_s\right)_{rq}\:.
\end{equation*}

\begin{figure*}[t]
 \centering
\includegraphics[width=0.45\textwidth]{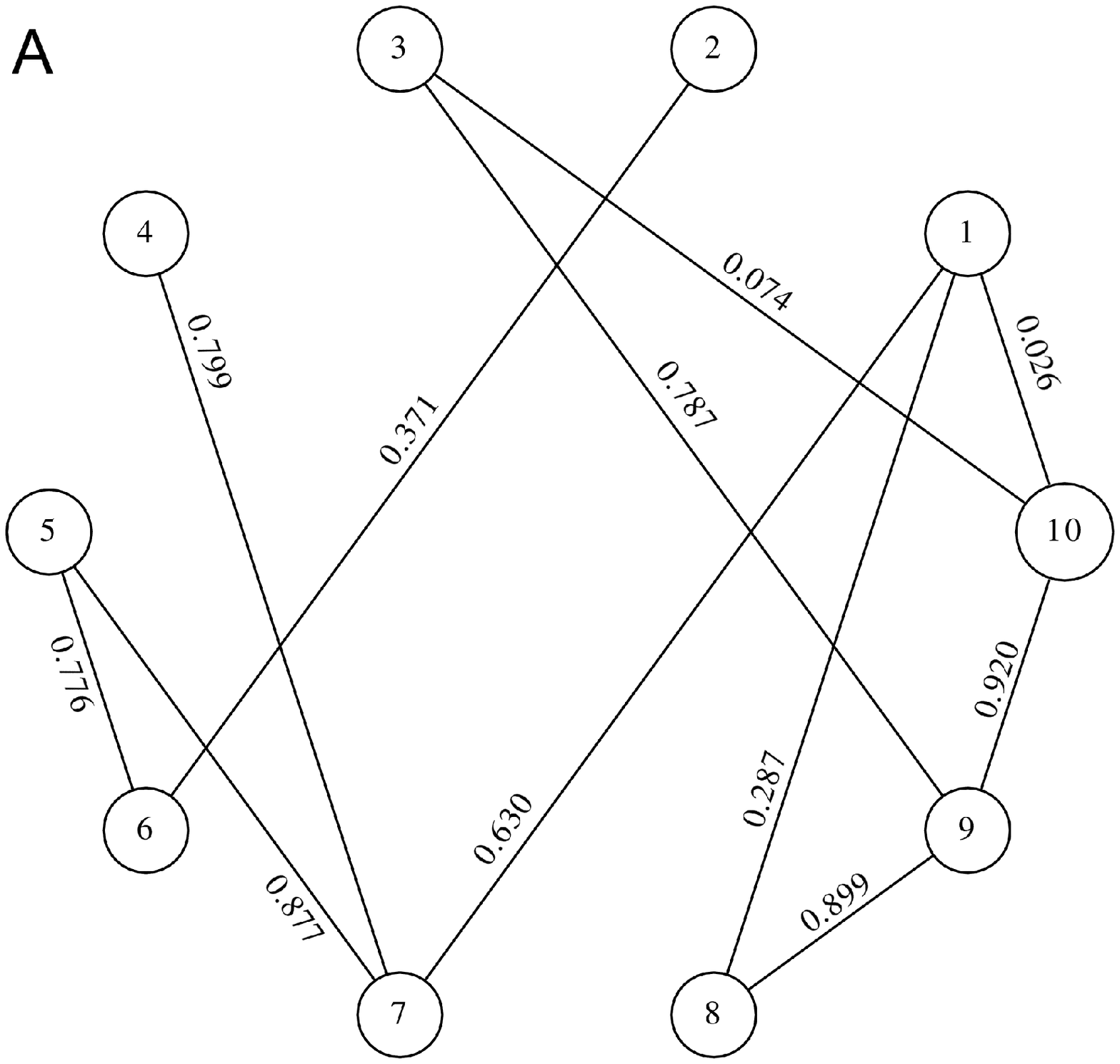}
\includegraphics[width=0.45\textwidth]{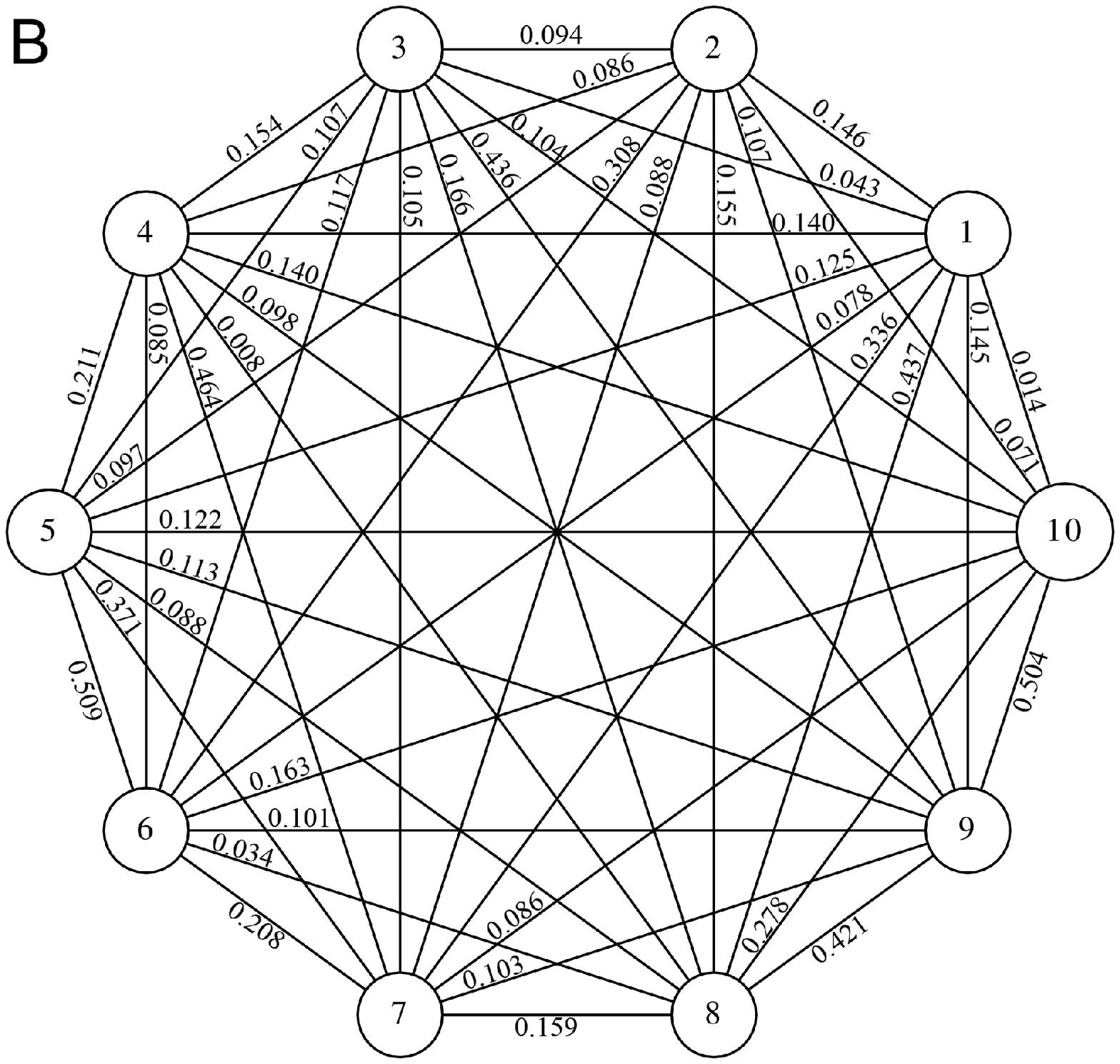}
\caption{\label{topotopotopo}Network topologies.
Each node is a chaotic Rössler oscillator. The edges
represent weighted diffusive connections. The weights
are indicated by labels in the vicinity of the corresponding
edges. The network switches between the configurations
in the two panels, remaining in each for some time.}
\end{figure*}
This allows a fully explicit expression for the boxed
term in Eqs.~\ref{variation} that accounts for the time
variation of the eigenvectors:
\begin{widetext}
\begin{equation}\label{NCbox1}
 -\sum_{i=1}^N\mathbf{v}_j^\mathrm{T}\left(t\right)\cdot\ddt\mathbf{v}_i\left(t\right)\boldsymbol \eta_i =
 -\frac{1}{t_1-t_0}\sum_{i=1}^N\sum_{k=1}^N\left[\sum_{r=1}^N\sum_{q=1}^N\sum_{x=1}^N R_{rk}\left(G_s\right)_{rq}R_{qx}\left(B_0\right)_{xj}\right]\left[\sum_{r=1}^N\sum_{q=1}^N\sum_{x=1}^N R_{rk}R_{qx}\left(B_0\right)_{xi}\dds\left(G_s\right)_{rq}\right]\boldsymbol\eta_i\:.
\end{equation}
\end{widetext}
However, for all practical purposes, one does not need
to use the expression above directly. In fact, considering
that most elements of $G_s$ are~0, it is quite simple
to compute and store $B_s$ in a symbolic form. Similarly,
most of the $\dds\left(G_s\right)_{rq}$ terms are~0. In
fact, they vanish if $r=1$, if $q=1$, if $\left|r-q\right|>1$,
if $r>2b+1$, and if $q>2b+1$, where $b$ is the number
of ``$\sin$--$\cos$'' blocks in $G_s$. Also, for all other
cases $\dds\left(G_s\right)_{rq}$ is proportional to a
sine or a cosine. Then, define $\dot{G_s}$ to be the matrix
whose $\left(rq\right)$ element is $\frac{1}{t_1-t_0}\dds\left(G_s\right)_{rq}$;
also, define $\dot{B_s}\equiv R^{\mathrm T}\dot{G_s}RB_0$.
Again, $\dot{B_s}$ can be easily computed and stored in
a symbolic form. Then, Eq.~\ref{NCbox1} becomes
\begin{equation}\label{NCred}
 -\sum_{i=1}^N\mathbf{v}_j^\mathrm{T}\left(t\right)\cdot\ddt\mathbf{v}_i\left(t\right)\boldsymbol \eta_i = -\sum_{i=1}^N\sum_{k=1}^N\left(B_s\right)_{kj}\left(\dot{B_s}\right)_{ki}\boldsymbol \eta_i\:.
\end{equation}
Once more, the equation above can be computed
symbolically, and evaluated at any particular
$t$, when needed.
Note that, despite the seemingly
complex expressions, Eqs.~\ref{NCbox1} and~\ref{NCred}
are straightforward to deal with. This is due
to the fact that $G_s$ and $\dot{G_s}$ are always
represented as tridiagonal matrices, and, as
mentioned above, many of the non-trivial elements
of $\dot{G_s}$ vanish as well. Thus, numerical
applications of this approach can benefit not
only from a restricted amount of needed memory,
but also from sparse matrix methods that result
in a small computational complexity.

The treatment we built is valid for every
positive, finite switching time $t^\ast\equiv t_1-t_0$
between configurations. As explained above, this time
cannot vanish, lest the derivatives in Eq.~\ref{variation}
be not defined. Nonetheless, one can wonder about the
behaviour of a system when the switching time becomes
very small, although non-zero. To this purpose, first
note that this time only appears as a multiplicative
factor in Eq.~\ref{NCbox1}. Thus, a very small $t^\ast$
would have the effect of making the boxed term in Eq.~\ref{variation}
much larger than the purely variational term. In this
regime, the effects on the the stability of the synchronous
solution are due mostly, if not exclusively, to the
switching process. In other words, if the expression
in Eq.~\ref{NCbox1} yields positive results, the synchronized
state is made more stable in the corresponding direction,
and vice versa for negative results, regardless of the
contribution coming from the variational term.
Note that this is in agreement with the finding that
blinking networks can greatly facilitate synchronization~\cite{Bel04,Sti06}.
Similarly, one can consider the opposite limit, namely
that of a secular switching for which $t^\ast$ becomes
very large, while still remaining finite. In this case,
for large enough switching time, the boxed term in Eq.~\ref{variation}
becomes negligible compared to the rest, and
the stability is determined entirely by the variational
term. This case is very similar to that of an evolution
along commutative structures~\cite{Boc06_2}. In fact,
in this regime of quasi-static evolution, the structure
at any given time $t$ is, to first order, equal to the
structure at time $t+\mathrm{dt}$. Thus, it is $\left[L_t,L_{t+\mathrm{dt}}\right]=\boldsymbol\varepsilon$.
Therefore, one can treat this case as the commutative
one with the addition of a small perturbation. Note that this
perturbation does not change the stability of the synchronized
state: for a positive variational term, instability
is maintained, and for a negative one, the synchrony
remains stable. The only uncertainty happens for
the critical condition corresponding to a vanishing variational
term, for which the perturbation can have either effect
on stability.
\begin{figure}[t]
 \centering
\includegraphics[width=0.45\textwidth]{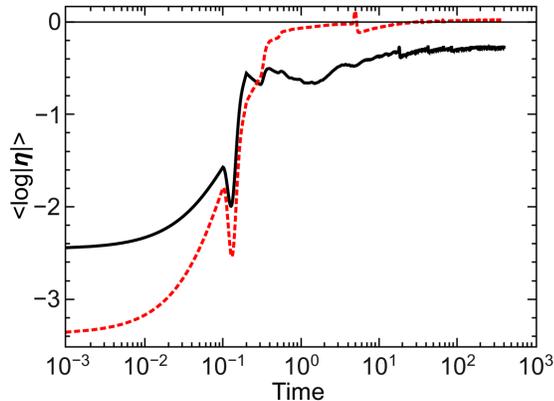}
\caption{\label{LyaEta}(Color online) Estimating the largest Lyapunov exponent.
The average of the logarithm of the norm of $\boldsymbol\eta$ (Eq.~\ref{etavec})
for the example system converges to approximately $-0.3$
when $\sigma=1$ (black solid line), and to approximately $0.022$ when $\sigma=0.1$
(red dashed line),
indicating that the synchronized state is stable in the first case,
and unstable in the second. Note the logarithmic time scale.}
\end{figure}

\begin{figure}[t!]
 \centering
\includegraphics[width=0.45\textwidth]{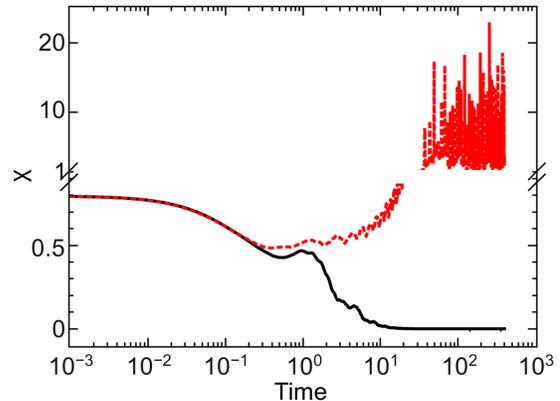}
\caption{\label{GlobErr}(Color online) Stability of the synchronized state.
The time evolution of the global synchronization error $\chi$
(Eq.~\ref{chierr}) shows that the system is able to reach
synchronization when $\sigma=1$ (solid black line), while it never
synchronizes when $\sigma=0.1$ (dashed red line), confirming the numerical results. Note the
logarithmic time scale and the break in the vertical axis.}
\end{figure}

To illustrate the use of our method, we consider the example
of a weighted network of $N=10$ chaotic Rössler oscillators,
switching back and forth between two topologies (Fig.~\ref{topotopotopo}).
Letting the
state vector $\mathbf x\equiv\left(x,y,z\right)$, each of the
oscillators obeys the local dynamics
\begin{equation}\label{numf}
 \mathbf f\left(\mathbf x\right) = \left(-y -z, x + 0.165y, 0.2 + z\left(x-10\right)\right)\:,
\end{equation}
with the output function
\begin{equation}\label{numh}
 \mathbf h\left(\mathbf x\right) = \left(0,y,0\right)\:.
\end{equation}
The switching times and the time periods for which the network
remains in each of the two configurations (permanence times) are
all set to $0.1$. We perform two simulations, one
with interaction strength $\sigma=1$ and one with $\sigma=0.1$,
to illustrate two different cases and the sensitivity of our method.
To estimate the largest Lyapunov exponent associated to the system
of Eq.~\ref{variation}, we compute the time-average of the logarithm
of the norm of the vector
\begin{equation}\label{etavec}
 \boldsymbol\eta\equiv\left(\boldsymbol\eta_2,\boldsymbol\eta_3,\dotsc,\boldsymbol\eta_N\right)
\end{equation}
at each integration step. The value to which $<\log\left|\boldsymbol\eta\right|>$
converges is $\Lambda_{\max}$. The results, in Fig.~\ref{LyaEta},
show that for the $\sigma=1$ case the convergence value
is approximately~$-0.3$, indicating that the synchronized state is stable.
Conversely, when $\sigma=0.1$, the estimated Lyapunov
exponent is just positive, with a value of approximately~$0.022$,
corresponding to an unstable synchronized state.
To verify this numerical result, we simulated the actual network
evolution for the two cases according to Eq.~\ref{eq1},
and with $\mathbf f$ and $\mathbf h$ given by Eqs.~\ref{numf} and~\ref{numh}
above. Figure~\ref{GlobErr}
shows the time evolution of the global synchronization error
\begin{equation}\label{chierr}
 \chi = \frac{1}{3\left(N-1\right)}\sum_{i=2}^N\left(\left|x_i-x_1\right|+\left|y_i-y_1\right|+\left|z_i-z_1\right|\right)\:.
\end{equation}
For the $\sigma=1$ case, after a certain transient, the synchronization error decays to~0.
When the interaction strength is lowered to $\sigma=0.1$,
instead, $\chi$ eventually starts growing and oscillates wildly,
always taking non-zero values. These results indicate that the system
is indeed able to synchronize in the first case, while it never
does in the second, in agreement with the numerical calculations
of $\Lambda_{\max}$. Thus, the simulations not only confirm the validity
of our treatment, but provide an example for which the stability
of the synchronized state can be changed by the tuning the parameters
controlling the topological evolution.

In summary, we demonstrated how to explicitly
solve the problem of constructing an appropriate
time evolution of a system of networked dynamical
units switching between different topologies.
Our method builds the evolution from a mapping
of the eigenvectors of the graph Laplacians of
the individual structures, and it ensures that
the elements of the eigenvectors are differentiable
at each intermediate time. This enables the use of the Master
Stability Function for network topologies that
evolve in time in a fully general and unconstrained
way.
While the connection
pathway is not unique, different solutions only
affect the variational part of the linearized
system.
It has to be remarked that our treatment
is valid \emph{regardless} of the time scales
involved. There is no restriction on the
permanence times of the network in each
configuration, and the only constraint
on the switching times is that they do not
vanish. In addition, our method
introduces a numerical advantage, in that
one only needs to integrate a set of linear
equations coupled with a single non-linear one,
rather than having to deal with a system entirely
composed of non-linear differential equations.
Also, this approach
does not rely on particular assumptions concerning
the structures visited by the systems,
and contains the regimes of blinking networks
and commutative evolution as its limiting cases.
Thus, our results have a natural
application in the study of synchronization
events in systems for which the temporal scales
of the topology evolution are comparable with
(or even secular with respect to) those characterizing
the evolution of the dynamics in each networked
unit. This is a common occurrence in many real-world
systems, such as neural networks, where synchronization
can become possible due to mutations~\cite{Noe03},
or financial market, where global properties
are affected by adaptive social dynamics.

\begin{acknowledgments}
The authors would like to thank Kyle Morris for helpful
discussions. CIDG acknowledges support by EINS, Network
of Excellence in Internet Science, via the European Commission's
FP7 under Communications Networks, Content and Technologies,
grant No.~288021.
\end{acknowledgments}


\begin{thebibliography}{99}
\bibitem{Alb02}   R. Albert and A.-L. Barabási, Rev.~Mod.~Phys.\ \textbf{74}, 47 (2002).
\bibitem{Boc06}   S. Boccaletti \emph{et al.}, Phys.\ Rep.\ \textbf{424}, 175 (2006).
\bibitem{Boc14}   S. Boccaletti \emph{et al.}, Phys.\ Rep.\ \textbf{544}, 1  (2014).
\bibitem{Var01}   F. Varela, J.-P. Lachaux, E. Rodriguez and J. Martinerie, Nat.\ Rev.\ Neurosci.\ \textbf{2}, 229 (2001).
\bibitem{Ber99}   E.~L. Berlow, Nature \textbf{398}, 330 (1999).
\bibitem{Boc02}   S. Boccaletti et al. Phys.\ Rep.\ \textbf{366}, 1 (2002).
\bibitem{Pec98}   L.~M. Pecora and T.~L. Carroll, Phys.\ Rev.\ Lett.\ \textbf{80}, 2109 (1998).
\bibitem{Lag00}   L.~F. Lago-Fernández, R. Huerta, F. Corbacho and J.~A. Sigüenza, Phys.\ Rev.\ Lett.\ \textbf{84}, 2758 (2000).
\bibitem{Bar02}   M. Barahona and L.~M. Pecora, Phys.\ Rev.\ Lett.\ \textbf{89}, 054101 (2002).
\bibitem{Nis03}   T. Nishikawa, A.~E. Motter, Y.-C. Lai and F.~C. Hoppensteadt, Phys.\ Rev.\ Lett.\ \textbf{91}, 014101 (2003).
\bibitem{Bel04}   I.~V. Belykh, V.~N. Belykh and M. Hasler, Physica D \textbf{195}, 188 (2004).
\bibitem{Cha05}   M. Chavez, D.-U. Hwang, A. Amann, H.~G.~E. Hentschel and S. Boccaletti, Phys.\ Rev.\ Lett.\ \textbf{94}, 218701 (2005).
\bibitem{Hwa05}   D.-U. Hwang, M. Chavez, A. Amann and S. Boccaletti, Phys.\ Rev.\ Lett.\ \textbf{94}, 138701 (2005).
\bibitem{Mot05}   A. Motter, C. Zhou and J. Kurths, EPL \textbf{69}, 335 (2005).
\bibitem{Yao09}   J. Yao, H.~O. Wang, Z.-H. Guan and W. Xu, Automatica \textbf{45}, 1721 (2009).
\bibitem{Li13}    F. Li and X. Lu, Neural Netw.\ \textbf{44}, 72 (2013).
\bibitem{Mas13}   N. Masuda, K. Klemm and V.~M. Eguíluz, Phys.\ Rev.\ Lett.\ \textbf{111}, 188701 (2013).
\bibitem{Den06}   R. Denamur and I. Matic, Mol.\ Microbiol.\ \textbf{60}, 820 (2006).
\bibitem{Csa07}   P. Csaba, M.~D. Maciá, A. Oliver, I. Schachar and A. Buckling, Nature \textbf{450}, 1079 (2007).
\bibitem{And88}   P.~W. Anderson, K. Arrow and D. Pines, \textit{The economy as an evolving complex system} (Addison-Wesley, Redwood City, California, 1988).
\bibitem{Han04}   J.-D.~J. Han et al. Nature \textbf{430}, 88 (2004).
\bibitem{Boc06_2} S. Boccaletti et al. Phys.\ Rev.~E \textbf{74}, 016102 (2006).
\bibitem{Amr06}   R.~E. Amritkar and C.-K. Hu, Chaos \textbf{16}, 015117 (2006).
\bibitem{Por06}   M. Porfiri, D.~J. Stilwell, E.~M. Bollt and J.~D. Skufca, Physica D \textbf{224}, 102 (2006).
\bibitem{Sti06}   D. Stilwell, E. Bollt and D. Roberson, SIAM J.~Appl.\ Dyn.\ Syst.\ \textbf{5}, 140 (2006).
\bibitem{Has13}   M. Hasler, V.~N. Belykh and I.~V. Belykh, SIAM J.~Appl.\ Dyn.\ Syst.\ \textbf{12}, 1007 (2013).
\bibitem{Has13_2} M. Hasler, V.~N. Belykh and I.~V. Belykh, SIAM J.~Appl.\ Dyn.\ Syst.\ \textbf{12}, 1031 (2013).
\bibitem{Jos08}   K. Josić and R. Rosenbaum, SIAM Rev. \textbf{50}, 570 (2008).
\bibitem{Bar02_2} L. Barreira and Ya. Pesin, \textit{Lyapunov exponents and smooth ergodic theory} (American Mathematical Society, Providence, RI, 2002).
\bibitem{Wu07}    C.~W. Wu, \textit{Synchronization in complex networks of nonlinear dynamical systems} (World Scientific, Singapore, 2007).
\bibitem{Sor11}   F. Sorrentino and M. Porfiri, EPL \textbf{93}, 50002 (2011).
\bibitem{War71}   F.~W. Warner, \textit{Foundations of differentiable manifolds and Lie groups} (Scott, Foresman and Company, Glenview, IL, 1971).
\bibitem{Noe03}   J.~L. Noebels, Annu.\ Rev.\ Neurosci.\ \textbf{26}, 599 (2003).
\end{thebibliography}
\end{document}